\documentclass{statsoc}
\usepackage{graphicx,float}
\usepackage{amssymb}
\usepackage{hyperref}

\title[Fair votes in practice]{Fair votes in practice}
\author[Denis Mollison]{Denis Mollison}
\address{Heriot-Watt University, Edinburgh, Scotland}
\email{denis.mollison@gmail.com}

\begin{document}
\vskip -40mm
\mbox{}\hfill{\em As submitted to RSS, 10 Feb 2023}
\vskip 40mm

  \begin{abstract}
\smallskip

Criteria for a good voting system have been given particularly careful scrutiny in recent years  \citep{BCfinal:2004,McAllister:2017,MVM:2019}, with general agreement that the core values are fair results, voter power and choice, and local representation.

Proportionality is a key aspect of achieving fair results.
It is a little over 200 years since the first election of a set of representatives under a proportional system, and 50 since the refinement of that system for the computer age.
This paper reexamines the basic ideas of that system, the `single transferable vote' (STV), which has voter empowerment and choice at its heart, and uses both core criteria and data from a range of political elections to evaluate its performance in comparison with alternative proportional representation (PR) systems in terms of both principles and practice.
The alternatives considered are List-PR and Mixed Member Proportional (MMP), also known as the Additional Member System (AMS).
\smallskip

I shall look particularly closely at proportionality, examining three aspects: broad linearity, the threshold for representation and the threshold for gaining a majority.
 
 \smallskip
As regards local representation,  an important question is how to design multi-member constituencies.
It will be argued that using constituencies based on natural demographic boundaries (such as local government areas) can combine better local representation with stability over time and better proportionality, 

\smallskip
The broad conclusions are that  List-PR is certainly an improvement on single-member pluraility, alias `First Past the Post' (FPTP), in proportionality, and has significant advantages on both the other criteria.
STV in turn improves on List-PR in respect of voter choice and local representation, and arguably in proportionality.
MMP, which is essentially a compromise between FPTP and List-PR, can be similarly proportional depending on how it is implemented, but has a range of significant potential problems, including non-linearity typically in the crucial near-majority case, and vulnerability to tactical voting \citep{Bochsler:2012}.

\smallskip
I shall make some use of the recent public availability of large preferential voting data sets for STV elections, which make possible new analyses of how STV functions in practice, and of voter behaviour, particularly examining how voters' second preferences relate to their first preferences.

 \end{abstract}
 

\medskip

\hrule

\section*{Contents}
\begin{enumerate}
\item[ ] Introduction
\item[1] Choosing a set of representatives
\item[2] Criteria
\item[3] Voter power and choice
\item[4] Local representation
\item[5] Proportionality and thresholds
\item[6] Patterns of preference
\item[7] Some problems and possible solutions
\item[8] Calculation and presentation of results
\item[9] Discussion
\end{enumerate}

\section*{Introduction}

This paper springs from an unashamedly positive motivation: to combine theory and data to help design the best possible electoral systems to support democracy.
Electoral reform is often pushed down the agenda because it does not directly affect social wellbeing; but the system for choosing political representatives is the foundation, the most basic building block of democracy.
If it is not proportional it gives more power to some parts of the population at the expense of others; and it has huge influence in determining the culture of politics (antagonistic or collaborative) and the ways in which it can change (adaptable or fossilised) \citep{Duverger:1964,Straw:2015}.
That reform is slow and often reversed should not surprise anyone familiar with elementary game theory: those who owe their power to an unfair system have a strong motivation to resist change.
The best remedy would seem to lie in better public understanding of its importance, and of the options for change.

This paper takes a fresh look at the main proportional voting systems, clarifying their raisons d'etre,  assessing them against `good system' criteria, and using election data to evaluate how they perform in practice.
The emphasis is thus on social principles rather than mathematical ones, and on data - what actually happens - rather than looking for theoretically possible anomalies.
There is a large literature on voting systems, ranging from politics to mathematical detail, but a good starting point for this study can be found in three mutually rather different approaches to identifying principles for a good electoral system: the British Columbia Citizens Assembly of 2004 \citep{BCfinal:2004},  and two recent reviews, one initiated by the Welsh parliament \citep{McAllister:2017}, and one by a grass-roots cross-party organisation \citep{MVM:2019}.
These two reviews came up with broadly similar criteria against which to assess voting systems (see Figure \S\ref{criteria_fig}); at their core, these revolve around the principles of fairness, voter power and choice, and local identifiability that were identified by the BC Citizens Assembly.

Key topics include primary questions: do all voters have equal power and sufficient choice? what is proportionality? how should we draw electoral boundaries? and a range of secondary questions aimed at improving our understanding of what happens in actual elections, such as what data on preferential voting can tell us about how voters view different parties.
Sections \ref{choice} to \ref{patterns} attempt to answer these questions, in each case using some new (though fairly simple) methodology.

I shall use data from a number of countries, primarily within Europe and particularly the British Isles - Ireland has been using the oldest proportional voting system, the {\em single transferable vote} (STV) for 100 years, while devolution within the UK has conveniently provided good data on a variety of systems over the past 25 years.
The largest of these data sets, for Scottish council elections, has provided for the first time a large body of complete preferential data that throw new light on STV.

In analysing the new riches of preferential data, it is tempting also to consider how best to make a single political choice, such as when electing a president or mayor, or in a referendum.
But I agree with \citet{Hill:1988} that this is a quite different matter from choosing a set of representatives; it will be the subject of a separate paper \citep{Mollison:2023}.

Supplementary material is online; also software, in particular a new R package for the computation, presentation and analysis of STV elections \citep{Mollison:2022}.

\section{Choosing a set of representatives}
\label{representatives}

The French Revolution led to some fresh thinking on how best to choose representatives, with Mirabeau in 1789 perhaps the first to state clearly the ideal of proportionality \citep{HoagHallett:1926}.
However, the first known election under a proportional scheme took place on 17 December 1819,  using an embryonic form of STV to elect a committee for the Birmingham Society for Literary and Scientific Improvement.

\subsection{The Single Transferable Vote (STV)}
\label{stv}

The basic idea of STV (in the US sometimes known by the simpler and perhaps better name of {\em ranked choice voting} or simply {\em choice voting}) is that proportionality should be achieved by matching the voters' preferences with the candidates in such a way that each person elected represents the same number of voters.
[We call the number of votes required for election the {\em quota}, $q$ say.]
The first scheme to implement this principle is due to Thomas Wright Hill (1763-1851), who ran a progressive school in Kidderminster, and he apparently had the original idea when organising the election of a representative committee from his pupils.

His idea was that the pupils should organise themselves in equal-sized groups, each of which could choose one representative.
This naturally leads to a process of several stages: initially the pupils form groups, according to their first preferences; then groups that are larger than necessary shed surplus pupils, who transfer to other groups, while members of groups that are too small disband to transfer to larger groups.
At the end of this process we should have a number of groups of exactly the required size $q$, each electing one representative, and a remainder of at most $q-1$ individuals unrepresented.

In formalising his scheme for the Literary and Scientific Society Hill introduced the idea of written votes, which needed to be revised in a series of rounds as transfers progressed \citep{Hill:1819}.
The key idea to make the scheme practical at large scale is preference voting, whereby each voter can in a single step completely specify how their vote should be treated.
The use of preference voting is due to Carl Andrae (1812-93), who when Prime Minister of Denmark independently invented a similar scheme to Hill's for the election of a new federal parliament, the Rigsraad, in 1855. 

Putting together the methodological ideas of {\em quota}, {\em transfers} and {\em preference voting}, we can give a precise algorithm for STV:

\begin{itemize}
\vspace{-2mm}

\item[(a)] votes are initially assigned to the voter's first choice;
\vspace{-2mm}

\item[(b)]
we calculate the number of votes a candidate needs to be elected (the quota);
\vspace{-2mm}

\item[(c)] if a candidate has more votes than needed, the excess is passed on by transferring the same proportion of each vote to that voter's next choice;
\vspace{-2mm}

\item[(d)] if not all seats are filled, the candidate with fewest votes is excluded, with the whole of each vote being transferred to the voter's next choice.
\end{itemize}
\vspace{-3mm}

\noindent
Steps (b-d) are repeated, as necessary, until all seats are filled.
\vskip 2mm

If this simple prescription is followed exactly, we have STV as defined by Meek (1969).
The reasons for the prevalence of a variety of untidier variants of STV will be explained in describing its history (\S\ref{history}).
\medskip

\noindent
{\em Notes:}

(a) To make the system precise we also need to specify the {\em quota} $q$; if there are $v$ votes and $s$ seats to be filled, then $q=v/(s+1)$ (since it is impossible for more than $s$ candidates to have more than this number of votes).\footnote{Andrae, and Thomas Hare (1806-91) writing two years later in England, suggested that the quota should be $v/s$;  in 1868 Henry Droop (1832-84) pointed out that $v/(s+1)$ suffices (his precise proposal was the integer value $[v/(s+1)]+1$).}
If as the count progresses some votes or parts of votes drop out (because they are to be transferred, but the voter has given no further preferences), then $v$ decreases, and $q$ can be reduced accordingly.

(b) One methodological detail so obvious that it might pass without being noticed is that each voter has exactly one vote, though it may be shared out among candidates through transfers.
This is where the name {\em single transferable vote} comes from.

(c) With Meek's definition of STV, the exact state of the count at any stage can be specified simply through the {\em keep values} $k_i$ of each candidate $i$, being the proportion that candidate keeps of each vote that reaches her.
The final keep values at the end of the count allow each voter to calculate exactly how their vote was used.

(d) Meek also suggested that voters should be allowed to give equal preferences to sets of candidates: equal preference for $n$ candidates involves (conceptually) splitting a vote into $m=n!$ micro-votes representing all possible orderings of those candidates, each with weight $1/m$.
This is impractical for hand-counting, but easy enough on a computer.

\subsection{History of STV}
\label{history}

The main difficulty in implementing STV lies in `transferring the same proportion of each vote' (step (c) above). The principle is easy to state (the first to do so was Gregory in 1881) but difficult to follow without a computer. As a result, over the period in which STV was being adopted for political elections, several slightly different STV methodologies evolved, varying in their compromises in fairness, clarity or repeatability made to facilitate counting votes by hand. In Ireland, the counting method still only deals in whole votes, so that surplus transfers are by a form of stratified sampling, meaning that results are not exactly repeatable \citep{Gallagher:1986}.

Meek (1969,1970) saw that the availability of computers allowed one to sweep away these compromises, and to follow exactly the simple  prescription of STV (as outlined in \S\ref{stv}).
And he saw that this would also achieve more strictly some of the basic advantages claimed for STV, including (a) treating voters equally (especially when making transfers of surpluses), (b) minimising wasted votes (therefore reducing the quota when any votes become non-transferable), and (c) minimising any incentive for tactical voting (the various hand-counting methods have unnecessary discontinuities that make possible a degree of tactical voting).
He also argued that voters being allowed to express their real preferences meant that they should be allowed to put some candidates equal, and (as mentioned in \S\ref{stv}) indicated how this should be done.

STV has been used continuously in Tasmania since 1909,  in Ireland since its first independent election in 1921, in Malta also since 1921, in Northern Ireland off-and-on but continuously for the Northern Ireland Assembly since 1998, and in Scotland for council elections since 2007.
It is also used for the Australian Senate, though in a form that makes it difficult to vote for individuals so that the system is close to a party list system.
With the conservatism typical of most countries as regards their voting systems, each has kept its own slightly different variant.

Meek's method was first used by a number of NGOs, including the Royal Statistical Society, in the 1980s, and for public elections in New Zealand in 2004; the version allowing equal preferences was first used by the John Muir Trust (for Trustee elections) in 1998, and by the London Mathematical Society in 1999.

\subsection{List Proportional Representation (List-PR)}
\label{list}

The second half of the 19th century saw the development of today's most used form of proportional representation, List-PR, which is based on parties rather than individuals.
Though later, this is a less sophisticated idea than STV,  corresponding to just the first stage of Hill's original scheme, {\em i.e.}  where pupils form their initial groups according to their first preferences.
Then the $i$th group is allotted a number of representatives $r_i$ equal to the size of the group, $v_i$ say, divided by a quota $q$, rounded to an  integer.

The pioneers of this development were aware that they were adopting a simplified alternative to STV; as one of their leaders, Ernest Naville, wrote in 1895: `I do not mean to say that the Swiss reformers considered the list
voting as the best in theory. I, for one, would prefer the method
which, like that of Mr. Hare, gives the elector a chance of
preferential vote without the party official list, for the purpose of
realizing better than any other the idea of representation' \citep{HoagHallett:1926}.

The commonest forms of List-PR are d'Hondt, using rounding down, so that $r_i$ is the integer part of $v_i/q$, and Saint-Lague, where $v_i/q$ is rounded to the nearest integer.
In each case, assuming as is almost universal that the total number of representatives $n$ to be elected is fixed,  $q$ is any number in the range of values for which the total of such allocations $\sum r_i$ is equal to $n$.
[For Saint-Lague, the system is sometimes modified so that the minimum number of votes required to get a representative is somewhat greater than $0.5q$; for example, Sweden has recently changed from using $0.7q$ to $0.6q$.]
There is no general formula for $q$; the commonest way of finding an appropriate value is by starting with the value, $v_{max}$, that would give $n=1$, and reducing it sequentially until the desired value of $n$ is reached.

The simplest form of List-PR is {\em Closed-list}, where the groups standing for election (usually political parties) put their candidates in an order that determines which of them will be elected.
Many countries ({\em e.g.} in Scandinavia), use {\em Open-list} systems where the voter can indicate preference for a specific individual candidate of the party of their choice.
 
 \subsection{Mixed Member Proportional (MMP/AMS)}
 \label{mmp}
 
A hybrid form of proportional representation was devised for West Germany during Allied occupation after WW2.
This is known as the Mixed Member Proportional (MMP) or Additional Member System (AMS), and is currently used in a number of countries, including Germany, New Zealand, Scotland and Wales.
It is usually described as a modification of the UK single-seat plurality system (FPTP), with `compensatory' List-PR seats added with the intention of making the overall result party-proportional.
But if the MMP system is working as intended, that is with sufficient compensation, the FPTP part makes no difference to the number of seats each party wins. It therefore offers more insight to think of MMP as a modification of List-PR; the distribution of seats among parties is determined by List-PR, but in deciding which individuals fill those seats priority is given to the winners of FPTP contests.

MMP can go `wrong' (meaning fail to achieve the same level of proportionality as List-PR) when the FPTP contests give a party more seats that it is entitled to by its list vote: this is known as an {\em overhang}.
The probability of overhangs occurring  depends on the proportion of compensatory seats, the size of regions (see next subsection), and the political demography of the country.
In the German parliament additional seats are added to compensate for overhangs, {\em i.e.} to restore overall party proportionality. However, in the Welsh and Scottish parliaments overhangs are simply accepted as part of the system, and have a significant effect on proportionality (see \S\ref{majority}).

\subsection{Districting and proportionality}
\label{districting}

With very few exceptions (notably Israel and the Netherlands), nations using proportional representation divide their country into regions or districts for electoral purposes.
For all proportional systems the size of these regions or districts has a significant effect on how they work, and indeed on their proportionality.
Districting will be discussed at length in \S\ref{local}, but some introductory remarks are appropriate here.

Small districts are good for local identifiability and for voters being able to assess individual candidates, but may give a less proportional result.
\citet{CareyHix:2011} analysed data from nearly 200 countries and concluded that using `low-magnitude' districts was associated with best governance.
Their explanation includes that `systems with median magnitude between four and eight seats \ldots offer opportunities for multiple winners, and thus afford voters an array of viable options, but at the same time do not encourage niche parties or overwhelm voters with a bewildering menu of alternatives'.
For STV, the slightly lower range of 3 to 7 seats has generally been found optimal, reflecting STV's superiority in achieving proportional outcomes in low-magnitude districts (see \S\ref{linearity}).

Note that the optimal range identified by Carey and Hix does not include the case of single-member districts.
Apart from their inability to provide proportionality, they have other defects.
For example, the phenomenon of safe seats whereby the same party holds a district for many years not only means that many voters feel disenfranchised, it also means that many potential candidates have no hope of being elected where they live: the result of this is `parachuting': a survey in 2017 by Demos found that less than one-third of Conservative MPs had a significant link to their constituencies prior to being adopted as candidates \citep{Morris:2017}.
It is then not surprising that a British Academy research report found that 
 ``The `constituency link' seems to be stronger under Open-List PR in Denmark or STV in Ireland than under the British FPTP system'' \citep{Hix:2010}.

There are two ways of districting,  which we may call `natural' and `equal': {\em natural} here refers to boundaries determined by demography, {\em equal} to where high priority is given to equality of {\em parity} (the number of electors per representative), and sometimes to having  equal numbers of representatives from each district.
Obvious concerns with using natural districting are that the variations allowed in parity and seat sizes might have an adverse effect on proportionality; but we shall see in {\S}s \ref{majority_stv}, \ref{equalisation} that, for  STV at least, the opposite is more likely to be the case.

\section{Criteria}
\label{criteria}

STV, List-PR and MMP were all conceived with the aim of empowering and giving fair representation to voters.
Against what criteria should we judge their success?

Possibly the most thorough evaluation of PR systems from the voter's point of view was that conducted by a Citizens Assembly in British Columbia in 2004 \citep{BCfinal:2004}.
This came down in favour of STV on the basis of three basic values, fair results, voter power and choice,  and local representation, together with the additional practical criterion of ease of use. More recently in the UK, the campaigning group Make Votes Matter came up with a set of 10 principles embodied in the Good Systems Agreement \citep{MVM:2019}; while in Wales a Commission set up by their parliament, the Senedd, came up with another set of 10 principles \citep{McAllister:2017}.

\begin{figure}
\includegraphics[width=\textwidth]{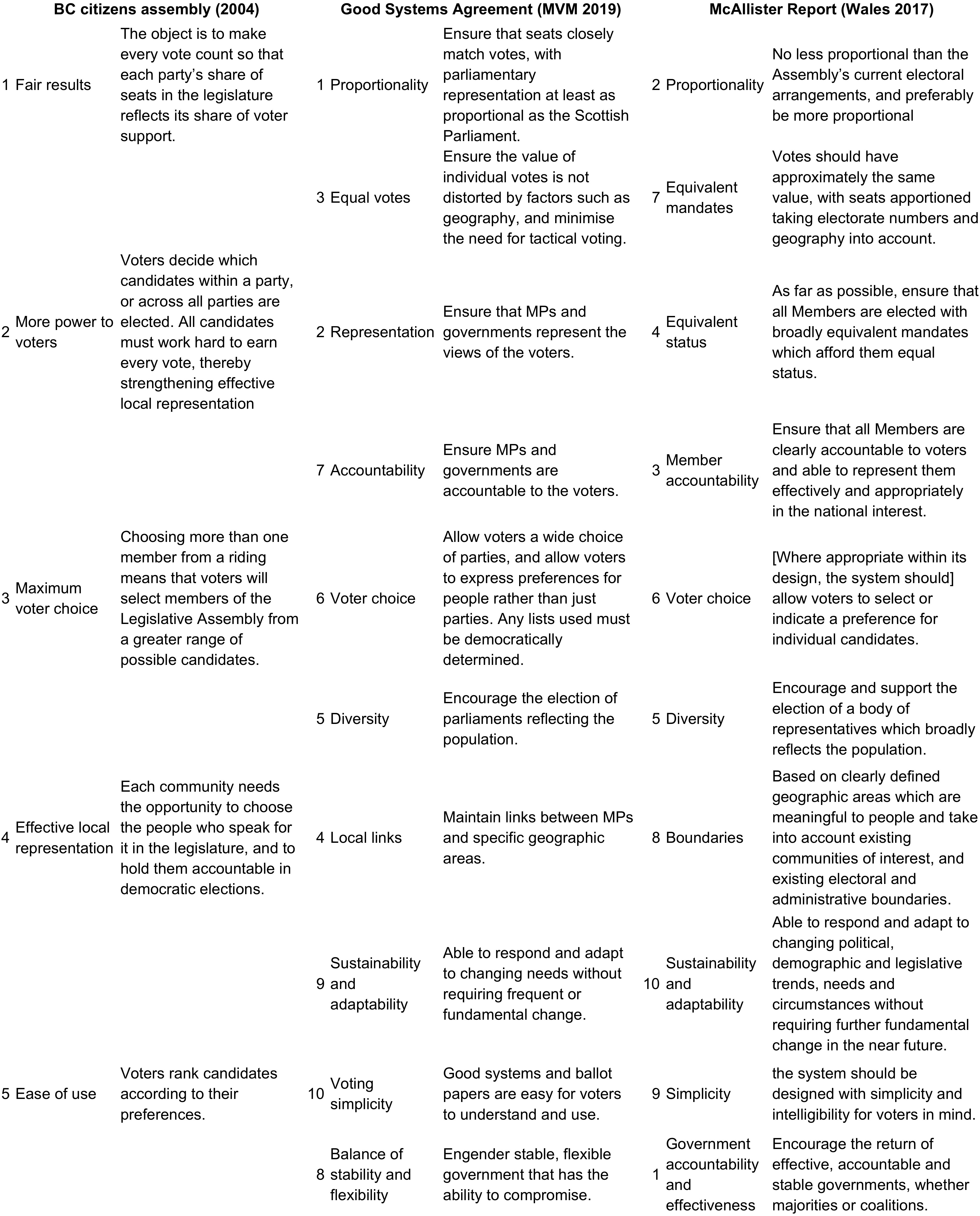}
\medskip
\caption{\label{criteria_fig}Criteria for a good voting system: comparison of BC 2004, GSA 2019, McAllister 2017}
\end{figure}

These latter two sets of principles are in very good agreement; though using rather different wording, they can be pretty well matched 1-1 as shown in Figure \ref{criteria_fig}. Most of them also match quite well with the rather simpler principles of the BC Assembly shown in the left column.
However, some of their wording relates not to fair voting so much as to good government, which is not covered in the BC criteria.  An extreme example of this top-down approach can be seen in the New Zealand Royal Commission report \citep{Wallace:1986}, which assumes that there is a good government model that the voting system should serve. This calls to mind \citet{Brecht:1953}'s sardonic comment:
`Would it not be easier in that case for the government to dissolve the people and elect another?'

The simpler BC criteria, in agreement with Thomas Wright Hill, reflect the alternative voter-centred, or bottom-up, approach, whereby government should (if necessary) be adapted to the representatives the voters choose, rather than {\em vice versa}.
There are persuasive arguments that voter empowerment through more proportional systems encourages more collaborative and flexible politics, as well as greater diversity among representatives. Certainly, FPTP has a strong tendency to result in two-party politics \citep{Duverger:1964} with the combativeness and inflexibility that implies \citep{Straw:2015}.
We shall look closely at the threshold vote required to win a majority ({\S}s \ref{majority}, \ref{majority}), but should bear in mind that constitutional reforms to encourage collaborative politics may make achieving a simple majority less crucial: the current power-sharing legislation of the Northern Ireland Peace Agreement \citep{NIPA:1998} provides a good example.

If we wish to examine how well specific voting systems satisfy the basic criteria of fairness, choice, and local representation, we need to see the system through the eyes of the voters: are they on an equal footing? can they express their real preference(s)? do they have one or more representatives they can identify with? This suggests simple practical criteria for evaluating a system:
\begin{itemize}
\item[(a)] fairness: votes are equally effective:
this implies that outcomes should be proportional in the sense that it takes at least approximately  the same number of votes to elect each representative
\item[(b)] voters have effective choices: this should include \\
(i) voters have a wide choice of candidates that have a reasonable chance of being elected, \\
(ii) voters can express their real preferences, and there is minimal incentive to vote tactically, {\em i.e.} other than in accord with those preferences,\\
(iii) voters can effectively influence whether their preferred candidates are elected, \\
(iv) voters understand the way in which their votes affects the outcome of the election, \\
(v) a minimal proportion of votes is wasted
\item[(c)] representatives have local ties - ideally voters should have a local representative that they voted for
\end{itemize}

\section{Voter power and choice}
\label{choice}

The three most-used proportional systems, as introduced in \S\ref{representatives}, are STV, List-PR and MMP.  All have good claims to satisfy the fairness criterion expressed in (a) above (but see discussions of proportionality in \S\ref{prop}). All can be implemented in a way that to some extent represents local ties - and for all this requirement has some conflict with achieving overall proportionality.
The chief respect in which they differ is voter empowerment and choice.

Here STV has an intrinsic advantage in that it is designed to make use of more information on the voter's preferences than simply their first choice.
Before comparing with alternative systems, it is worth while critiquing some variants of STV, flourishing chiefly in Australia, that undermine that intrinsic advantage.
First, filling casual vacancies by `countback' of the previous election's votes results in many more candidates, and pressure to allow what are effectively party lists.
If this is compounded by a rule that voters must give a minimum number of preferences - or even that they must put {\em all} candidates in order - it is not surprising that voters mainly choose the party list option, even when this is further compounded by parties being allowed to control their full preference order, which offers wide scope for tactical vote management by parties \citep{Raue:2022,Difford:2022}.
Such modifications not only comprehensively undermine the idea of STV and the good system criteria (b) above; they also (repeating \citet{CareyHix:2011}'s words) `encourage niche parties and overwhelm voters with a bewildering menu of alternatives'.

\subsection{List-PR}
\label{list_pr}

The advantages of STV over List-PR are perhaps most clearly seen if we consider how one might modify the latter so as to better meet the criteria set out in (b) above.
First, some versions of List-PR (`Open-list') allow voters to express preference for individual candidates; STV takes this further by allowing the voter's preference order to range across all candidates, not just those of one party. [Allowing equal preferences as can be done in STV should satisfy those who do wish to simply vote for a party.]

Second, if a candidate has more votes than they need, STV allows the surplus votes to be transferred; and when excluding a candidate allows their votes to transfer to a candidate still in the contest.
This elimination of wasted votes has the technical benefit of setting a simple quota for election: if you are electing $n$ representatives, they are elected if and only if they have more than a $1/(n+1)$ share of the votes.

But the crucial benefits of allowing transfers are that it avoids wasted votes, and that it allows voters to express their real preferences (for example, for a candidate from a minor party or an independent) without fearing that their vote will be wasted if that candidate is excluded.
Without transfers, it is not only that a party can more easily achieve a majority because there are wasted votes that are ignored; worse, the vote for a party that perhaps justifies a majority may be based on tactical voting and thus not reflect their true support.
Finally, a system with transfers such as STV encourages a more cooperative politics, with parties benefitting from transfers from others with similar policies.

If we look at the comparison the other way round, it is interesting to note that STV can be turned into d'Hondt List-PR by imposing restrictions: not allowing transfers of any kind, only allowing voters a first preference, which has to be an equal preference for the candidates of one party, and by letting that party determine how the resulting tied results among their own candidates shall be decided.

\subsection{MMP}
\label{mmp2}

List-PR in turn has many advantages over MMP.
The idea of MMP is that it embeds FPTP style local representation in a List-PR proportional system.
Unfortunately, in doing so it combines most of the faults of both systems. In particular, it has safe seats of both types -  FPTP seats in localities where one party dominates, and safe list seats by virtue of being hghly ranked in a party list. It adds further problems for the voter through the interaction of the two components. If MMP is working as intended, the winner in a constituency simply displaces a winner of the same party on the list.
This sets up conflicts both of fact - constituency candidates are in competition with others of the same party on the list, but it is difficult to assess who might displace whom, and therefore for voters to vote on the merits of those candidates - and of perception: although the constituency results should make no difference to party seat totals, they are seen as the chief battleground between parties.

Constituency results can matter when there are overhangs; as already mentioned (\S\ref{mmp}), these disrupt the proportionality of the system.
In UK elections using MMP, such overhangs are allowed to stand, with the other parties getting fewer seats than their  list entitlement; in New Zealand both overhangs and list seat entitlements stand, so that the total number of representatives is increased by the size of the overhang. In Germany, overhangs and list entitlements both stand, but additional `compensation' seats are added to restore proportionality.
We shall look more closely at the occurrence of overhangs, and the associated disproportionality, in \S\ref{majority}.

In Germany the number of compensatory seats needed may be large. In 2017 the number of members had to be increased from 598 to 709, and while that restored party proportionality at both state and national level, it introduced a disproportionality in parity, with the number of members per voter varying by up to 20\% between states; to have rectified that disproportionality as well would have required increasing the parliament to 1001.

Finally, MMP is open to tactical voting.
On the list there are incentives not to vote for a party if it is likely either to be below threshold or to achieve overhangs.
This can achieve innocent distortions, as with the informal alliance of the pro-independent SNP and Green parties in Scotland, whereby the SNP win more constituencies, and the Greens more list seats, than they might otherwise have done, with a mutual benefit arising from the SNP overhangs that result.
More seriously, it is possible to game the system by setting up `decoy' parties to gain similar benefits.
In Scotland a new pro-independence Alba party made this their explicit intention at the 2021 election, but failed to get above the threshold to win any list seats.

However in several countries that do or did use MMP this `decoy' party tactic has been very successful \citep{Bochsler:2012}.
In Albania in 2005 the two main alliances each set up decoy parties, with the effect of 
more than halving the representation of minor parties.
And under a similar system in Italy, the same tactic by the alliance led by Berlusconi in 2001 was so successful that they won more list seats than they had candidates to fill.

It is perhaps not surprising that the MMP system is poorly understood: in a survey of voters with experience of the system in Scotland, over 50\%  when surveyed gave the wrong answer to basic binary questions on how it operates \citep{Bromleyetal:2006}.
Overall, MMP performs badly on all the criteria of (b) above, List-PR being better at least for (b)(ii-iv), and is the only one of the three systems compared to have serious problems with sustainability and stability (because of natural drift into overhang territory as observed in Germany (see \S\ref{majority}), and the possibility of tactical cheating as in Albania and Italy), or with the McAllister Report's criterion of `Equivalent status' (see Figure \ref{criteria_fig}).

\section{Local representation}
\label{local}

As already mentioned, there is evidence from PR systems around the world that they work best in multi-member constituencies - `districts' for short - electing a small number (between 3 and 8) of representatives \citep{CareyHix:2011}.
Some countries (for example Denmark and Norway) have formal constitutional constraints, requiring administrative districts to have separate representation.
But if we have freedom to choose electoral district boundaries, how should we choose?

The traditional approach to districting has been to give primacy to parity ({\em i.e.} the number of electors per representative), but to allow some flexibility to improve the fit to natural communities.
The Venice Commission, without giving any specific reason, recommends that parity should not vary beyond $\pm$10\%, or 
$\pm$15\% in `special geographical circumstances'; some countries, notably the US, insist on much tighter parity (though at the same time allowing party-controlled gerrymandering that has far greater distorting effects!).

Another traditional strand, especially for STV, has been to insist on `equal-size', {\em i.e.} that each district has the same number of representatives; some go further in insisting that this must be an odd number ({\em e.g.} the Proportional Representation Society of Australia, following \citet{Howatt:1958}).
Tasmania, Malta and Northern Ireland all use STV with 5-member districts.

Here I shall follow through the implications of instead giving priority to fitting natural communities, and not worrying about parity or equal-size.
What variability in parity does this lead to? and more importantly, what variability from overall proportionality?
The reason for taking this approach is that we have already identified that voters having a link to particular `local' representatives is a key feature of a good electoral system; and in practice voters' concerns over districting show much more concern over boundaries than over parity.
Inherent in the attraction of localism is that the localities involved should be meaningful, preferably defined in terms of natural communities, and that their role as electoral districts should be stable over time.

I shall concentrate here on districting for STV because while it is generally agreed that it needs a districting process, the details, especially the question of variable or fixed size, have for long been a key topic of debate.
List-PR and MMP have often been used within already defined regions of widely varying size, so that the districting question hardly arises; but with increasing interest in good local representation it has recently received more attention, at least within the UK \citep{McAllister:2017}.
Much of what I shall say here applies equally to List-PR and MMP, though the sizes of their districts may need to be rather larger to attain reasonable proportionality. 

Since natural communities vary unsystematically in their numbers of electors, a scheme that fits them well must allow variability in size, the number of elected members for a district, as well as in parity.
Since they also vary in topology, there are no certain rules for joining up or splitting communities to form contiguous districts.
However it is found in practice that a twofold range, {\em i.e.} `any size between $x$ and $2x$', where $x$ is chosen to give a number of seats in a desired range, makes the process of districting feasible, but that a smaller range does not.
The intuitive reason for this is that a twofold range guarantees that combining two districts that are too small cannot give one that is too big,
nor can halving one that is too big give two that are too small.

The requirement for a twofold range gives two options within the 3-8 seat `sweet spot',
either 3-5 seats or 4-7: the former gives more emphasis to localism, the latter to proportionality.
Districting then becomes a matter of choosing areas solely on the basis of their best fit to communities within an appropriate twofold range; for the 3-5 seat case this could be $2.7<e<5.4$, where $e$ is the area's entitlement calculated as $e=E \times S/N$ ($E=$ district electorate, $N=$ total electorate, $S=$ total seats).

A major advantage of giving natural boundaries priority in this way is that they do not need to change over time. Changes in electoral numbers can instead usually be accommodated by changing the allocation of seats between districts; if an allocation rule is agreed ({\em e.g.} the US's Huntington-Hill proportionally-nearest integer) this updating can be done instantaneously whenever electoral numbers are updated.
The price paid for local identity and stability is some variability in parity: the worst case is where a district's entitlement falls half-way between two possible integer sizes, which can mean a variation from parity around the Venice Commission maximum of  $\pm 15\%$.
We shall see in \S\ref{equalisation} that the effect of such variability on overall proportionality is smaller than one might expect.

That the natural communities method is feasible and achieves stability in practice can be confirmed from examples. McAllister (2017) includes a scheme for the Senedd (the Welsh parliament) of 90 members in 4-7 member constituencies based on local authority boundaries. If this had been adopted at the creation of the Senedd in 1999, it could have been used with no change of boundaries since then,  just a small number of $\pm$1 changes  in seat numbers, and with standard deviation in parity steady at about 6\%.

A similar scheme for the UK is illustrated in Figure \ref{stv_map} - for full details see the Supplementary material.

\begin{figure}
\centerline{ \includegraphics[width=0.98\textwidth]{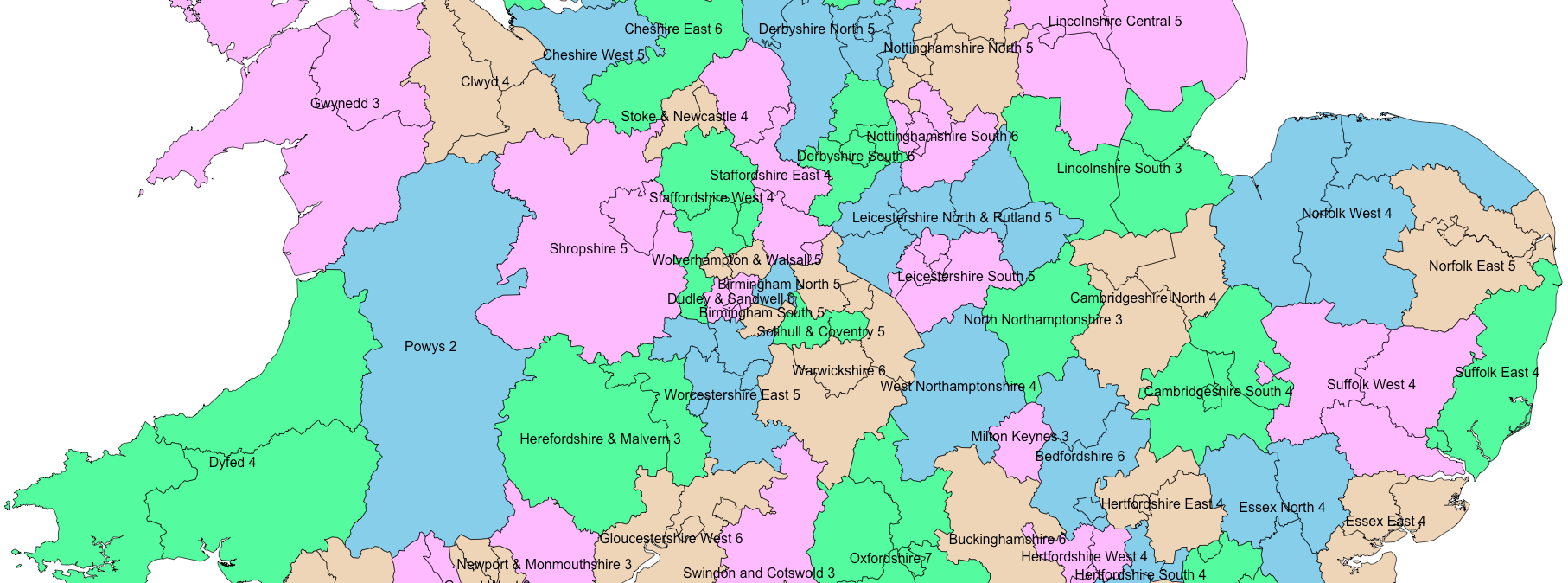}}
\caption{\label{stv_map}Part of the map of an STV scheme for the UK parliament based on Local Authority areas. Each coloured area represents a constituency.}
\end{figure}

\subsection{Isolated and sparse communities}
\label{minorities}

A common problem in territories with diverse geography is whether and how to accommodate isolated or sparse communities.
The BC Citizens Assembly (2004), while recommending up to 7-seat districts for populous areas, proposed a small number of 2-seat districts for the remoter rural regions.
This is a flexibility to be used with restraint, as wards with fewer than 3 seats have an adverse effect on overall proportionality.
More recently, the Electoral Reform Act (Scotland) 2020 introduced greater flexibility for council wards, extending the range from 3-4 to 2-5, with the possibility of 1-seat wards in the case of islands.
The first review under this Act, in 2020-21, is interesting both in demonstrating how the flexibility of allowing 2-seat wards can give a much better fit in sparse and island areas (particularly in this case in Eileanan--an-Iar), and also how achieving a good fit to communities naturally leads to an agnostic view of parity (see Supplementary material).

Although the Islands Review was carried out with the overt aim of demarcating wards with parity appropriate to 2, 3, 4 or 5 seats, the outcome is a distribution of seat entitlements that is not significantly different from uniform over a range from below 2 to above 5, just as though the `natural communities' method proposed here had been used.
The review also made allowance for special geographic circumstances for sparse areas; while no numeric amount is specified, their recommendations correspond closely to adding 10\% to such areas' entitlements.
It is interesting to see that there have been no public objections to the larger than usual variations in parity, nor to the extra allowance for sparse areas in the recommendations.
To the contrary, the review's proposals for Highland Council were rejected by the Scottish Parliament because of strong feeling throughout that council that {\em not enough} allowance was made for sparse areas.

More radically, there is a good argument for departing far from parity when a parliament is representing a confederation or a collection of independent territories of very different sizes.
\citet{Penrose:1946}, thinking of the newly founded United Nations, showed that if such components vote independently for or against each issue, then for them to have influence on decisions proportional to their populations they should have representation proportional to the square root of those populations (see also \citet{Kurz_etal:2017}).
In this context it is interesting to note that the representation of countries in the European Parliament is closely fitted by a 2/3 power law\footnote{See Supplementary Material}.

\section{Proportionality and thresholds}
\label{prop}

Proportionality, the idea that $x$\% of votes should give $x$\% of seats, seems a very simple idea, but is much less so when considered carefully.
Numerically, proportionality is often assessed using a single statistic, such as the standard deviation of seats from votes (in \%s) \citep{Gallagher:1991}.
We shall assess proportionality in more detail, in three aspects: broad linearity and two key thresholds:

(a) Dependence of seats on votes over the main range of votes, from any threshold up to the level at which a party can expect an overall majority: for proportionality, this relationship should be linear, with slope close to 1 and little variability from the expected value.

(b) Threshold for representation, or more broadly the relation between votes and seats for minor parties.
It is widely thought that a threshold level of votes to win representation is a good thing, discouraging too much political fragmentation. Relevant questions are: how large is the (explicit or implicit) threshold? Is it sharp or smooth? what happens to the votes cast for below-threshold parties or candidates? and how does it affect proportionality?

A basic problem with thresholds, shared by List-PR and MMP, is that votes for minor parties may be depressed by tactical voting, so that the \% of votes for each party does not truly reflect the voters' preferences.
As noted in \S\ref{mmp}, under MMP there can be additional distortion where supporters of major parties vote tactically for another party on the list because they expect their preferred party to win all or more than its share of seats in constituencies (a potentially self-fulfilling prophecy).

(c) Threshold for a majority:  a crucial question is what \% of votes can be expected to give a party or coalition 50\% of seats? This should be close to 50\%, and again, there should be low variability about this threshold value.

\begin{figure}
\centerline{ \includegraphics[width=0.98\textwidth]{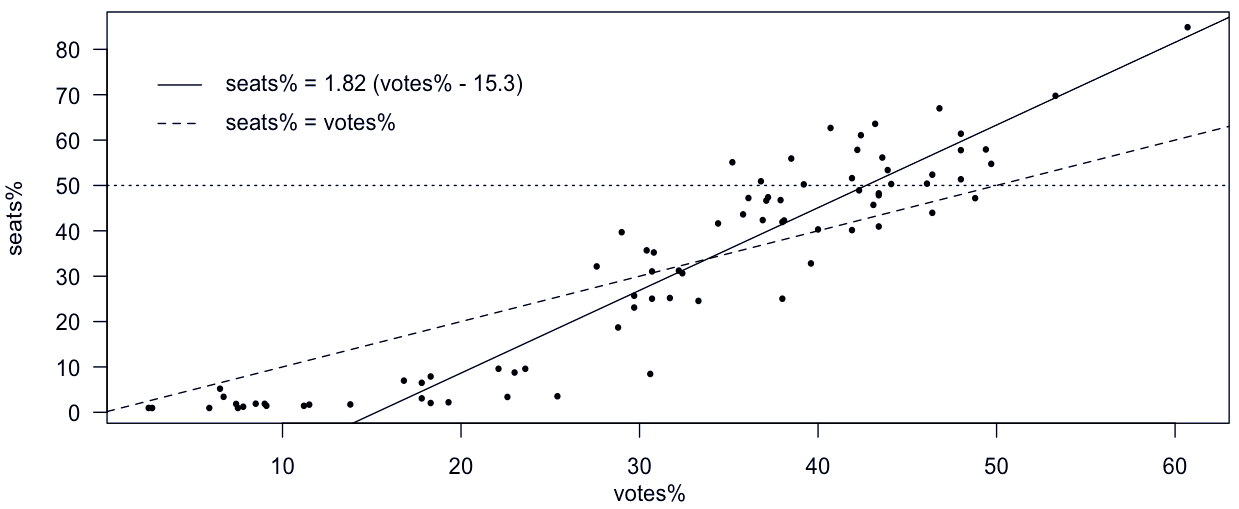}}
\caption{\label{vs_fptp}Relation between national vote and seat \%s for UK general elections 1922-2019}
\end{figure}

To illustrate these aspects, let us look first at just how disproportional is the FPTP system (plurality in single-member seats). Figure \ref{vs_fptp} shows the relation between votes and seats for the 27 UK General Elections of the past 100 years.

On average, over the range of votes from 15 to 60\%, a party has a seat:vote ratio of 1.82, {\em i.e.} it gets 1.82\% seats for each extra 1\% of votes (solid line in figure). If this disproportionality were consistent, there would be an argument in its favour as allowing a clear winner with a mandate to govern, but FPTP is not consistent: there is huge variability about the average ($\pm$8.5\%). Two examples of this variability are the 1983 election, when two parties with 27.6 and 25.4\% of votes got respectively 32.3 and 3.5\% of seats; and that of 2015, when parties with 12.9 and 4.7\% of votes got respectively 0.2 and 8.6\% of seats.

As to threshold, only the geographically concentrated Nationalist and Northern Irish parties have ever won significant numbers of seats with less than 15\% of the national vote. And as to majority, the linear fit estimates that 42.7\% of votes is required to give a majority of seats, but because of the great variability a majority has been achieved with 35.2\% of votes (in 2005), and not achieved with 48.8\% (in 1951).

Before looking at relevant data for proportional systems, we should note that STV does not fit easily into the straightjacket of most proportional voting analysis, which looks only at (implicitly first choice) votes for parties.
It is therefore important to stress that the idea of STV, matching representatives to equal numbers of voters by allowing transfers, is in a sense (within each district) exact proportionality.
And, because preferences are expressed for individual candidates rather than simply parties, STV gives proportionality in terms of whatever is most important to the voter, which might be cross-party issues.
[This is why it is especially good for non-political elections, where the outcome should reflect the proportions of voters who support different priorities of an organisation ({\em e.g.} conservation, campaigning or education for an environmental charity).]

A key advantage of STV is that it avoids wasted votes (whether surplus or excluded) through its use of transfers. This has qualitative benefits, for voters who can express their real preferences, knowing that if their first preference does not have sufficient support their vote will not be wasted but passed on; and for parties in encouraging more collaborative politics and campaigning. But it also has quantitative benefits, in that it comes closer to requiring 50\% support to win 50\% of seats (see \S\ref{majority_stv}).

\subsection{Linearity}
\label{linearity}

Figure \ref{prop} shows generalized linear model (GLM) fits for data from a range of parliamentary and assembly elections.
These examples are chosen from systems having a strong element of local representation. Thus List-PR  is represented by the GB elections for the European Parliament (1999-2019), and estimated outcomes from Northern Ireland Assembly elections to 6-seat constituencies (1998-2016) - the latter allows an interesting comparison with the actual STV results for those elections.

\begin{figure}
\centerline{ \includegraphics[width=0.98\textwidth]{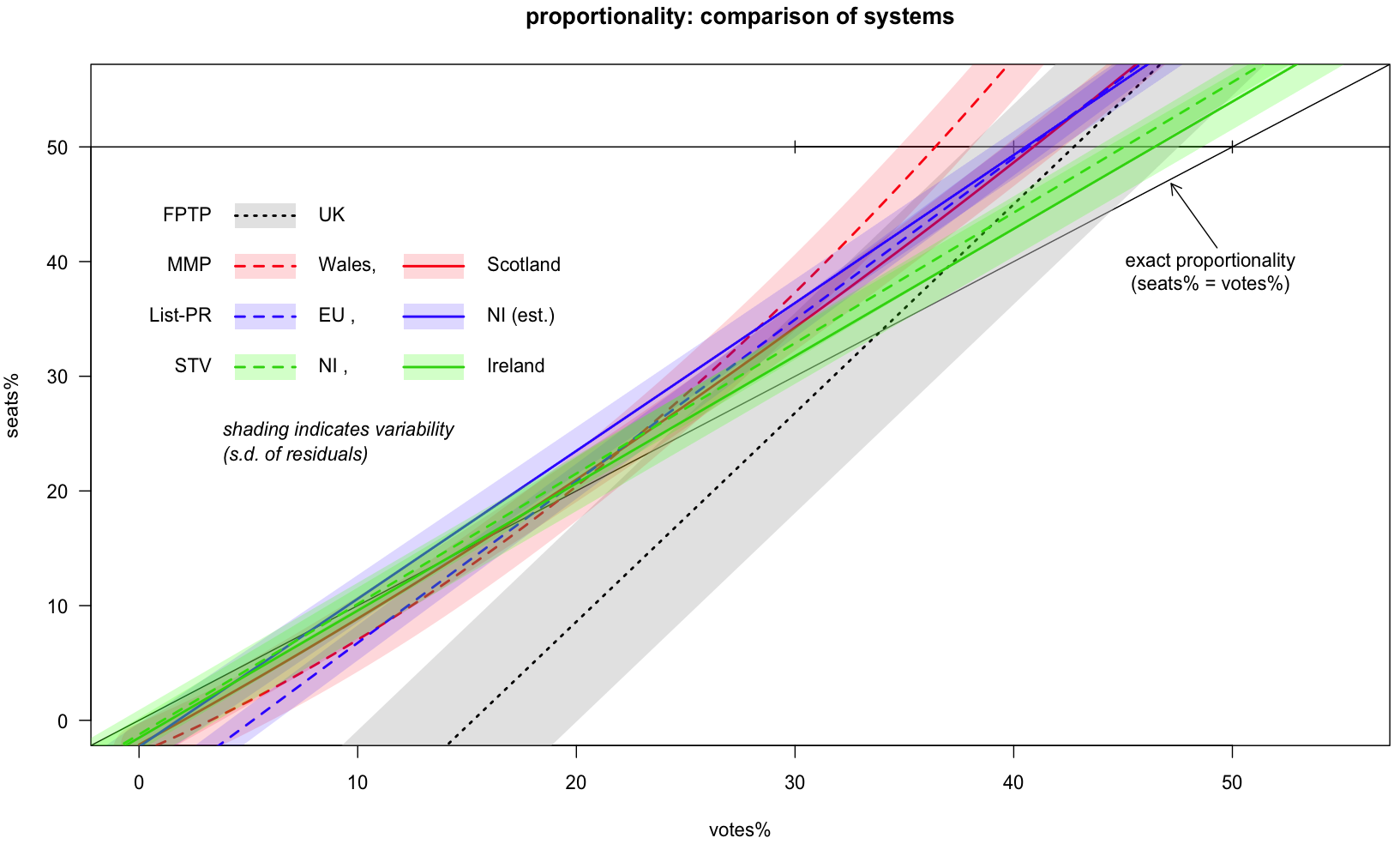}}
\noindent
\caption{\label{prop_comparison}Proportionality: comparison of systems used in the UK and Ireland}
\end{figure}

It is perhaps surprising that STV achieves the best proportionality among these examples, despite the data used being first preferences rather than simply votes for a party.

It is further interesting to note that the smaller variable-sized (3-5) seat system of Ireland has better proportionality than the 6-seat system of Northern Ireland (1998-2016) - a point I shall return to in \S\ref{majority_stv}.
The variability of the Irish system is somewhat higher (standard deviation of residuals 2.0\% as against 1.2\%), which is at least partly explained by transfers being relatively more important in smaller size districts.
Perhaps surprisingly, the List-PR examples have similar variability (1.4 and 1.8 \% respectively).

The fitted slope parameters (`seat:vote ratios') for the two STV examples are 1.10 (Ireland 1951-2019) and 1.14 (Northern Ireland 1998-2016), and for the two List-PR examples are 1.41 (GB EU elections, 3-11 seat districts) and 1.29 (estimates using NI Assembly 6-seat districts).

Can we explain the linear relationships we have found, particularly for List-PR?
For a single region or district, vote data can be well-fitted using the simple approximation that the total vote for parties not gaining seats (`wasted votes') is a fraction $w$, and each party gaining seats,  by achieving some multiple of a quota $q$, has a surplus $x$ uniformly distributed on [0,q) for d'Hondt, and on [-q/2,q/2) for Saint-Lague; it also needs to be noted that the probability of being the party that wins the last seat, and thus sets the quota, is roughly proportional to that party's vote.
This leads to an estimate for Saint-Lague, $s_i = v_i/(1-w)$, {\em i.e.}\ exact proportionality if wasted votes are ignored.

\begin{figure}
\centerline{ \includegraphics[width=0.98\textwidth]{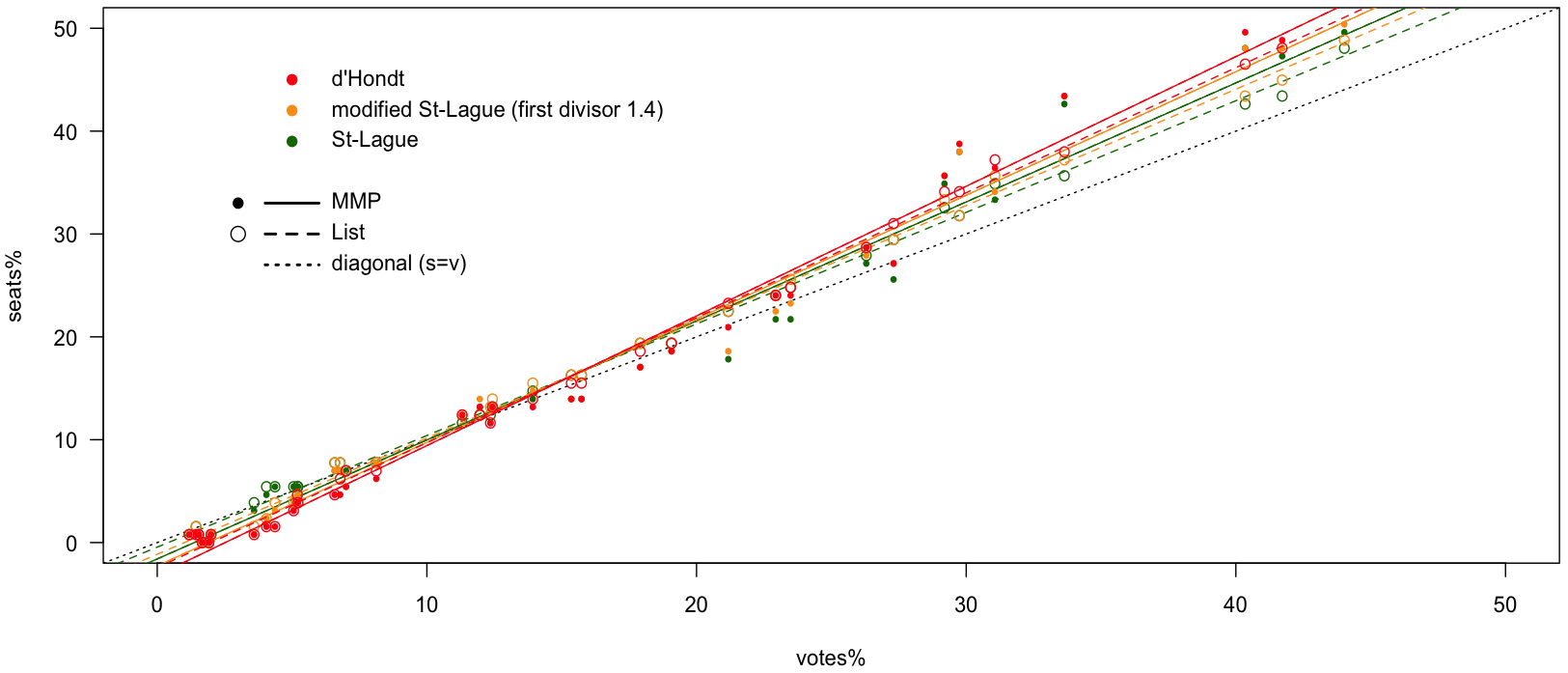}}
\caption{\label{SL_DH}Effect of varying both voting system (MMP or List) and allocation system (St-Lague, modified St-Lague or d'Hondt), using Scottish Parliament data (1999-2021).}
\end{figure}

In contrast, the d'Hondt rounding down method favours larger parties, and a similar approximate calculation gives $s_i=b(v_i-v_0)$, where $b=(1+m/2n)/(1-w)$, where $m$ is the number of above-threshold parties and $n$ the number of seats in the region; and $v_0 = 100/2nb$.
Thus for a region with 16 seats, 4 main parties, and with 10\% wasted votes, we expect $s_i=1.25(v_i-3.9) \%$, compared with $s_i=1.11 v_i$ for Saint-Lague.
If, as is likely, the proportion of wasted votes falls under St-Lague, its proportionality will be even better.
For example, using Scottish Parliament list vote data (1999-2021) we find overall national fits of $s_i=1.22(v_i-2.0) \%$ for d'Hondt (with 8.9\% wasted), $s_i=1.09(v_i-0.4) \%$ for St-Lague (with 6.8\% wasted) (see Figure \ref{SL_DH}).

In regions with fewer seats (and thus closer to the Carey-Hix ideal) the seat:vote ratio will be significantly higher, as for the UK's EU Parliament elections (3-11 seats) or in estimates for 6-seat districts using the first-preference data for Northern Ireland Assembly elections.
For example, for the latter at district level the seat:vote ratio is around 1.5, though as we have seen it is only 1.29 at the national level, owing to demographic variability between districts.
This reduction can be less marked for larger districts, for example if we compare the list part of Scottish or Welsh parliamentary votes at regional and national levels, so  it is not predictable whether having a smaller number of larger districts will necessarily give better overall proportionality than a large number of small ones.

In contrast to both List-PR and STV, the MMP examples show significant deviations from linearity (see Figure \ref{wales_prop}).
Because the overhangs responsible for this occur principally when one party is close to a majority, this will be discussed in the majority-threshold section below (\S\ref{majority}).
If we ignore the non-linearity, linear fits to these data are comparable to the List-PR examples (1.45 for Wales, 1.27 for Scotland).

\subsection{Threshold for representation}
\label{thresh}

\begin{figure}
\centerline{ \includegraphics[width=0.98\textwidth]{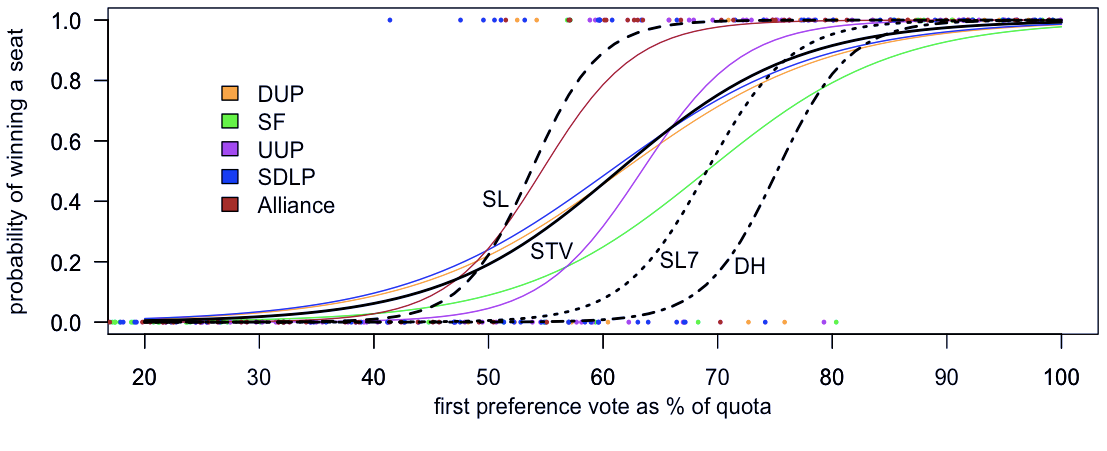}}
\caption{\label{thresh_ni}Threshold comparison for List-PR (d'Hondt, St-Lague and modified St-Lague) with STV (with, in colour, fits for individual parties for STV), using Northern Ireland Assembly data (1998-2016).}
\end{figure}

For any electoral system based on multi-member constituencies, there is an implicit threshold for the vote percentage required to win seats. The quota for election under STV is $q=1/(s+1)$ ({\em e.g.} 20\% in a 4-member constituency), but this of course does not have to be all in first-preference votes.
The relation between first preference votes and probability of election is shown in Figure \ref{thresh_ni} for the main parties in Northern Ireland Assembly elections: the vote level required to achieve 50\% probability of election varies from 55 to 69\% of a quota,  with the Alliance Party and Sinn Fein at opposite extremes, reflecting their different abilities to attract transfers.
Figure \ref{thresh_ni} also shows the thresholds for election if we apply List-PR rules to the STV first preference data.
D'Hondt gives a higher threshold, unsurprisingly as in party terms it resembles STV without transfers (see \S\ref{choice}).
Saint-Lague gives a lower threshold, similar to the best-performing party under STV, but does not make that discrimination between parties that attract transfers and those that don't.
Modified Saint-Lague, with threshold 0.7 instead of 0.5 of a quota ({\em i.e.} first divisor 1.4 rather than 1), is closer to d'Hondt.

Similar figures for the Scottish Council elections of 2012, 2017 and 2022 show similar numerical results.
Their larger data sets allow us to look at how the various party thresholds have changed over these three elections.
The main changes have been that Labour and the Liberal Democrats now have rather lower thresholds, probably due to exchanging more preferences in a realignment following the Scottish independence referendum of 2014; while the Greens now have a higher threshold, having aligned themselves as a pro-independence party\footnote{See Supplementary material}.

\subsection{Threshold for a majority}
\label{majority}

In both examples of small-district List-PR we find that the threshold for a majority is close to 41\%.
This could be improved, {\em i.e.} made closer to 50\%, by either increasing district size, or, as in countries such as Denmark and Sweden allocating a number of `top-up' seats at national level to make the result as close as possible to exact proportionality.
However, both of these modifications diminish the quality of local representation, and because they generally give nothing to parties below some threshold (in Sweden 4\%), a party or coalition can still win a majority on less than 50\% of the national vote.
In Germany in 2013 the compensatory system allocated the CDU/CSU alliance 49.7\% of seats, exactly matching their 49.7\% of above-threshold list votes, but their actual overall vote was only 41.5\%.

\begin{figure}
\centerline{\includegraphics[width=0.98\textwidth]{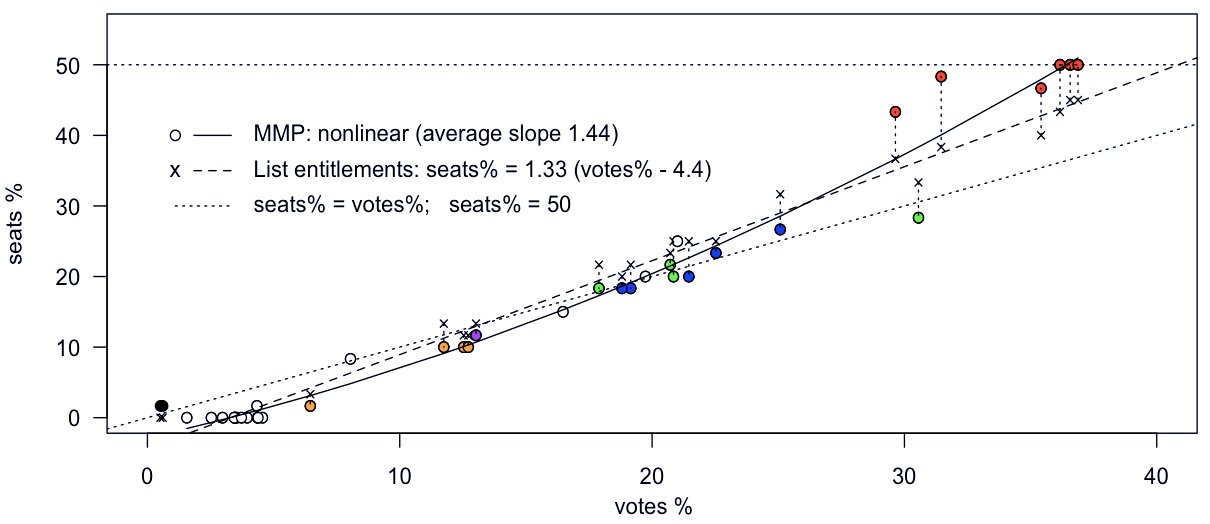}}
\caption{\label{prop_wales}MMP: Welsh Senedd 1999-2021, showing how overhangs introduce nonlinearity}
\end{figure}

For MMP, if the system is working as intended, the numbers of seats won are determined by the list votes, so that proportionality is as above.
However, MMP can have overhangs, where a party wins so many constituency FPTP contests that it gets more seats than the proportional list system would entitle it to.
This has been discussed in general terms in \S\ref{choice}.
In numerical terms, where overhangs are accepted as for the Scottish and Welsh parliaments, the effect is significant in the vote range that determines whether a party can win a majority.
Figure \ref{prop_wales} shows the effect of overhangs in Welsh Senedd elections: the dependence of seats on votes is significantly quadratic rather than linear, and the estimated threshold vote required to win 50\% of seats is reduced from 40.8\% to 36.4\% - as is confirmed by the three elections (2003, '11, '21) in which one party has indeed won exactly 50\% of the seats on between 36 and 37\% of the votes.

The frequent occurrence of overhangs in Wales is not surprising, as the List:FPTP seat ratio is only 50\%.
However, even in Germany with its List:FPTP seat ratio of 1:1, substantial overhangs can occur: in their most recent elections (2017, 2021) their top-up to achieve proportionality at national level has required adding over 100 seats to their 598-seat parliament.
This plethora of overhangs can be explained by Germany's mostly politically homogeneous regions, and to there being a larger number of parties with substantial support so that constituencies can be won on a low vote.
Indeed, the number of effective parties in Germany, as measured by \citet{LaaksoTaagepera:1979}'s index, has increased from 3.2 in 2002 to 5.8 in 2021.
The number of overhangs has much increased in Germany in recent years, supporting the reasonable conjecture that under a proportional system the effective number of parties tends to increase, which can in turn be expected to increase the probability and size of overhangs.

\begin{figure}[h]
\centerline{ \includegraphics[width=0.9\textwidth]{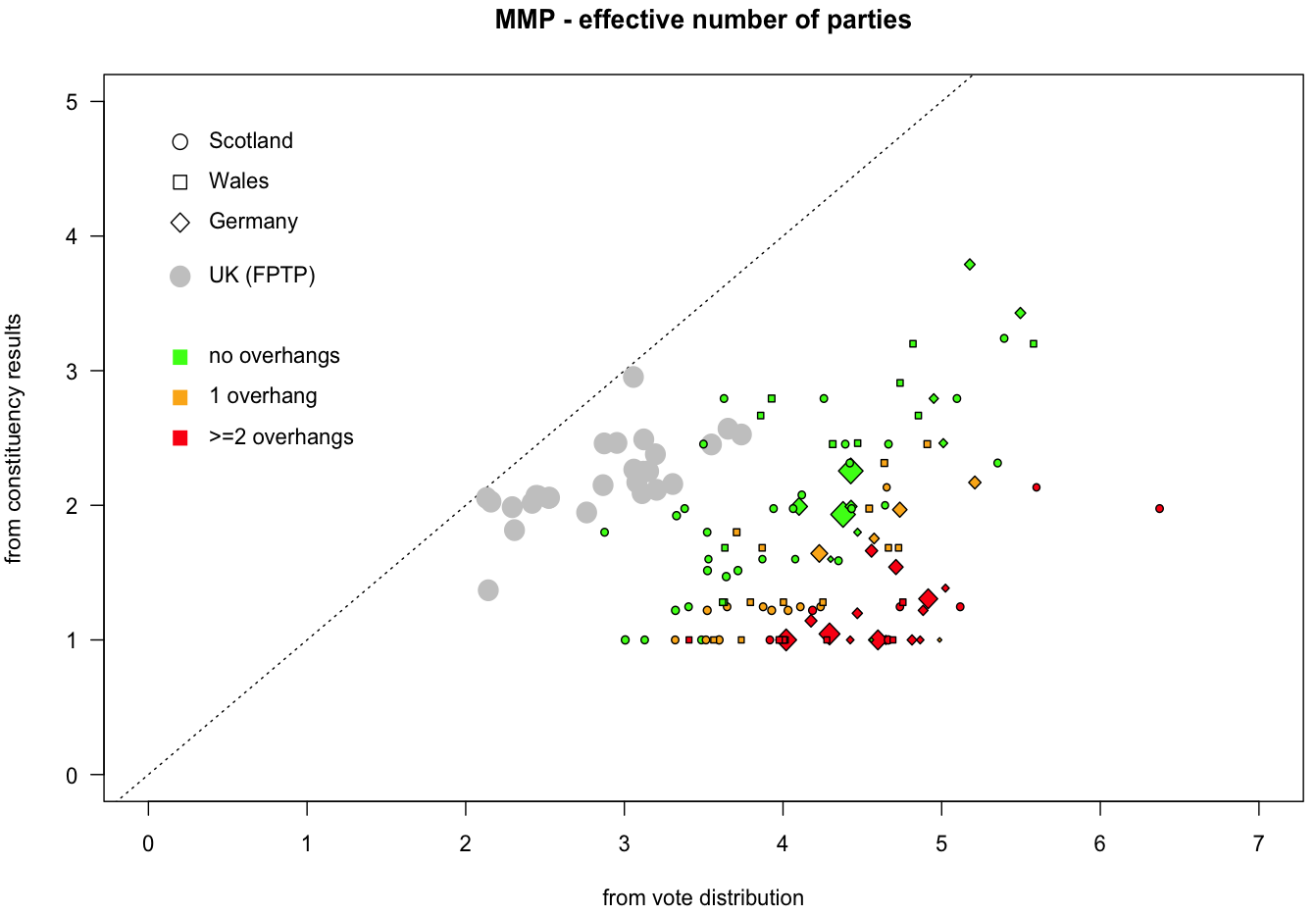}}
\caption{\label{nparties}The relation between the number of parties as calculated from votes with the number calculated from constituency seats, for parliaments using MMP with regions. The area of each symbol is proportional to the number of seats in the region. The same relation for UK Parliament elections using FPTP is shown for comparison.}
\end{figure}

Figure \ref{nparties} illustrates the association between the larger effective number of parties, as calculated by the Laakso-Taagepera index, and the occurrence of overhangs. 

\subsection{Winning a majority under STV}
\label{majority_stv}

How does STV perform when a party or coalition are close to 50\% of the vote?
In what is  arguably the most important case, where voters are divided between two clear alternatives (Left  vs.\  Right or Leave vs.\  Remain, for example), a simple argument suggests that it is highly accurate. Here each side may be represented by a mixture of parties (and possibly independent candidates), but we assume all transfers are within-side.
Then the outcome in each district (provided there are sufficient candidates from each side) will be determined simply by the number of quotas of first preferences on each side: a side with proportion $v$ of the vote in a district with $s$ seats will win $[(s+1)v]$ seats.
More generally, the outcome in each district depends on what we may call the final preference distribution, {\em i.e.} the distribution of votes at the final stage of the count, not that of first preferences; but if the first and final preference distributions differ this indicates that there are transfers between the sides, so that thinking of the election as simply `L v R' is not entirely correct.

For an example of how this might work out in practice, possibly the best available set of data is from the UK 2016 referendum, where voting figures for Remain vs Leave are available for each council area, and thus for the STV districting scheme of Figure \ref{stv_map}.
This is an interesting empirical distribution, in that it is significantly skew.
We adjust the votes very slightly (under logit transformation) to give a distribution where each side has 50\%.
We can not only calculate how many seats each side would have won  (49.8:50.2\%) but can also, by randomising the sequence of district sizes relative to the empirical vote distribution, calculate the mean and standard deviation for such a scheme (a roughly equal mix of mainly 3-6  seat districts) for the kind of demographical vote variation represented by the referendum data: we find 49.9 $\pm$ 0.5 \%. Thus the conclusion is that the majority-threshold for STV with this kind of demographic vote distribution is very close to exact proportionality, certainly much closer than any List or MMP system is likely to achieve because of their omission of wasted votes.

\begin{figure}
\centerline{\includegraphics[width=0.98\textwidth]{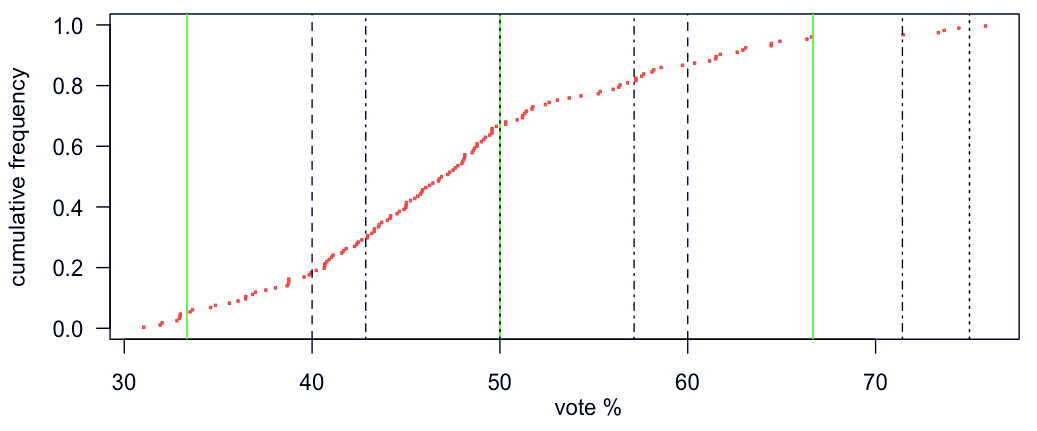}}
\noindent
\caption{\label{brexit}Distribution of Remain vote in 2016 UK referendum, showing STV quota cutoffs for 3 (.....), 4 (---), 5 (green) and 6 (-.-.-.) member seats}
\end{figure}

This data analysis can be taken further to look at the difference between variable and equal-size districts.
We can use the same randomising idea to estimate the majority threshold if we used all 5-seat districts, as in Northern Ireland, Tasmania and Malta. This gives a seat total estimate with the same variability ($\pm$ 0.5 \%) but with a mean (49.2\%) that is significantly different from 50\%.
The reason variable district sizes do better is apparent if we plot the empirical vote distribution together with the cut-offs for the main district sizes used - see Figure \ref{brexit}.
What's going on is that we're trying to estimate a continuous distribution using either one or several rather broad discretisations: it's not surprising that the `several' alternative does better at capturing the skewness of the vote distribution.
Indeed, under the 5-seat option almost all districts return results of either 3-2 or 2-3, with the dividing line therefore drawn at the median (here 48.7\%)  rather than the mean (50\%) of the distribution, as it should be for proportionality of votes to seats.

\subsection{Equalisation: the effects of variation in parity}
\label{equalisation}

The question of how variations in parity affect overall proportionality was raised in \S\ref{districting}.
However, `equalisation', {\em i..e.} a strong emphasis on minimising variation in parity (as in the current UK Parliament boundary revision), requires a very detailed process, and is therefore typically updated only at long intervals (typically 10 years or more).
Over such a period, the pattern of demographic change can introduce significant bias between party totals, for example if there are significant population movements between urban and rural areas \citep{Baston:2010}.

In contrast, the natural districts approach described in \S\ref{local} allows seat allocations to be kept up to date, thus not giving time for such bias to develop.
We can use the technique of \S\ref{majority_stv} to estimate the effect of variability in parity for this approach.
The effect of such variability is to give a party $n_i$ seats in a district, where its fair entitlement would be the non-integer value $m_i=n_i e_i/s_i$, where $e_i/s_i$ is the parity ratio for the district.
If we again use the 2016 referendum data as typical of demographic variability in voting,
and the STV scheme of Figure \ref{stv_map}, we find that the effect of parity variability is $\pm 0.2\%$ ($\pm$ 1.3 seats in a 650 member parliament).
The explanation for this  perhaps surprisingly small effect in relation to the $\pm 6\%$ variation in parity at district level is that it is the difference in seats between the two political sides in each district that counts, not the total number of seats.

An important note is that we are assuming in this subsection, both for  equalisation and for natural districts, that boundaries are drawn up by an independent body, {\em i.e.} that there is no gerrymandering.

\section{Patterns of preference}
\label{patterns}

The uniquely large set of full preference data available for the Scottish Council elections (two councils for 2007, all councils for 2012-2022, a total of nearly 1100 data sets) provides a rich seam for exploration of patterns of political preference.
In this section I shall just scratch the surface with three examples.

\begin{figure}
\centerline{\includegraphics[width=0.65\textwidth]{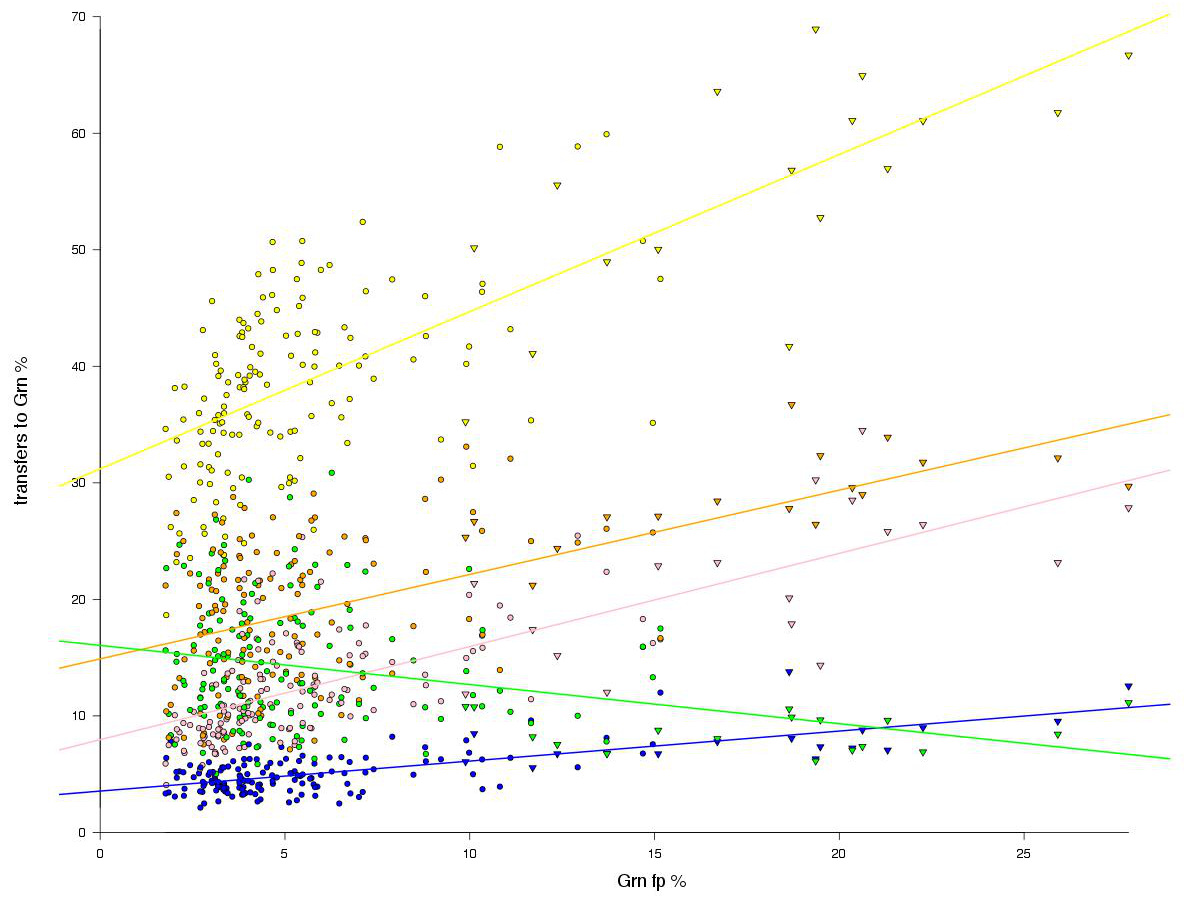}}
\noindent
\caption{\label{prefs2_sc}Scottish Council election 2017:  How the \% of transfers going to Green candidates from other main parties depends on the Green first preference vote.
Note that the green points and line show the \% of transfers from Green going to non-transferable.}
\end{figure}

First, it is common to assume that transfers between parties can be represented by a simple matrix, whose entries $t_{ij}$ represent the probability that a supporter of party $i$ will have next preference party $j$.
An alternative - a sort of null hypothesis - would be that second preferences are independent of first preferences $v_j$, and have the same distribution.
Linear regression shows that the truth lies between these two extremes.
As illustrated in Figure \ref{prefs2_sc}, the transfer frequencies show a roughly linear dependence on first preferences of the transferee, $t_{ij} = a_{ij}+b_{ij}v_j$, where $0<b{ij}<1$.

\begin{figure}[h]
\centerline{\includegraphics[width=0.35\textwidth]{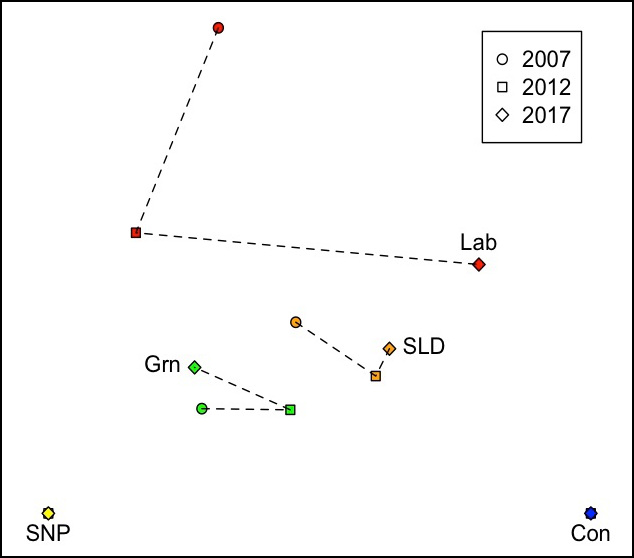}}
\noindent
\caption{\label{dparties}Scottish Council elections 2012-22:  MDS plot showing changes in closeness of voters' perceptions of main parties.}
\end{figure}

One simple way of using transfer preferences is to define a measure of distance between parties (as perceived by voters).
These can then be viewed in two dimensions using non-metric multi-dimensional scaling \citep{JardineSibson:1971}:
Figure \ref{dparties} shows how such distances between the main parties have changed over the three Scottish Council elections from 2012 to 2022.
Starting from approximate equidistance from the SNP and Conservatives, Labour moved to being perceived as closer to the SNP during the latter's first period in national government, but then to being closer to the Conservatives after the 2014 Independence referendum.
The Greens and Liberal Democrats both moved closer to the Conservatives in the first period, but then diverged as Scottish politics became more polarised between pro-independence and unionist parties.

Lastly, the detailed Council election data can be used to explore the patterns of preference in each election, for example to examine the frequency of occurrence of Condorcet cycles, where, if `$>$' means `is preferred to', $A>B>C>A$.
This is of theoretical interest because most if not all examples of pathological behaviour in STV rely on data with cyclic preferences.
We find that Condorcet cycles occur in less than 5\% of elections, and almost all involve 2 or more candidates of one party; this is explored further in \href{supp/condorcet.pdf}{\citet{Mollison:2023}}, which also goes into
the dependence of second preferences on the distribution of first preferences in greater depth.

\section{Some problems and possible solutions}
\label{problems}

One of the advantages of STV is that where a party stands more than one candidate it is the voters' preferences that decide which of them (if any) is elected.
Three problems have arisen in this context, which give rise to varying degrees of unfairness as between candidates, voters and parties.

First, there is `donkey voting', where voters through laziness or indifference rank the candidates in the order that they appear on the ballot paper. 
Voting analysis suggest that around 15-20\% of voters behave like this, leading to substantial unfairness between candidates of the same party.
There is also, surprisingly, a significant ballot-order effect between candidates of different parties, even between parties as different as the Conservatives and the Scottish National Party: that is, in wards where each of these parties has one candidate, there is a significant tendency to give first preference to whichever party appears first on the ballot paper.

The logically best answer to this problem is to vary the order of candidates on the ballot paper.
A full solution would involve randomly permuting the order on each paper.
However, the partial solution of using just one order, preferably random, and alternating it with its reverse, would  be much simpler to implement and should greatly reduce the problem.

Second, a significant number (around 1\%) of ballot papers have to be disallowed because the voter has placed the same number, or an `x', against each of the party's candidates.

Lastly, both in order to mitigate donkey voting, and to promote less well-known candidates, parties often go in for a form of vote management in which, if they have 2 candidates A and B, they ask voters in one half of the ward to vote `AB', and in the other half to vote `BA'.

It would avoid or much reduce these latter two problems if voters were allowed to express equal preference between candidates.  STV allowing equal preferences is perfectly possible, and has been in use for elections of trustees in some member organisations in the UK for 20 years.

\section{Calculation and presentation of results}
\label{presentation}

{\subsection{Software}
\label{software}

Traditional hand-counting of votes is well known to suffer from errors: for example, the count in the UK's North East Fife constituency in 2017 required 3 recounts, in each of which the winner changed.
Physical ballots plus electronic counting is arguably best: it can be audited by checking physical batches of ballots against corresponding computer data; and allowing access to the electronic vote files so that anyone interested can do their own check of the officially calculated result.

All STV methods are straightforward to program.
Meek STV, which as already mentioned needs electronic counting, is at the same time more elegant and robust; it is conceptually simpler, avoiding arbitrary choices and messy programming.
New software is discussed in the Supplementary material, and \href{https://github.com/denismollison/stv/blob/master/stv_pkg_manual.pdf}{can be found on Github} \citep{Mollison:2022}.

{\subsection{Presentation of results}
\label{results}

Clear presentation of results is important for public engagement and acceptance.
The software cited above \citep{Mollison:2022} produces graphs for each stage of an STV count ({\em e.g.} Figure \ref{stv_stage}), showing where any transferred votes originated, for both Meek STV (with or without equal preferences) and the Weighted Inclusive Gregory (WIG) system used for Scottish Council elections.

\begin{figure}[h]
\centerline{\includegraphics[width=0.98\textwidth]{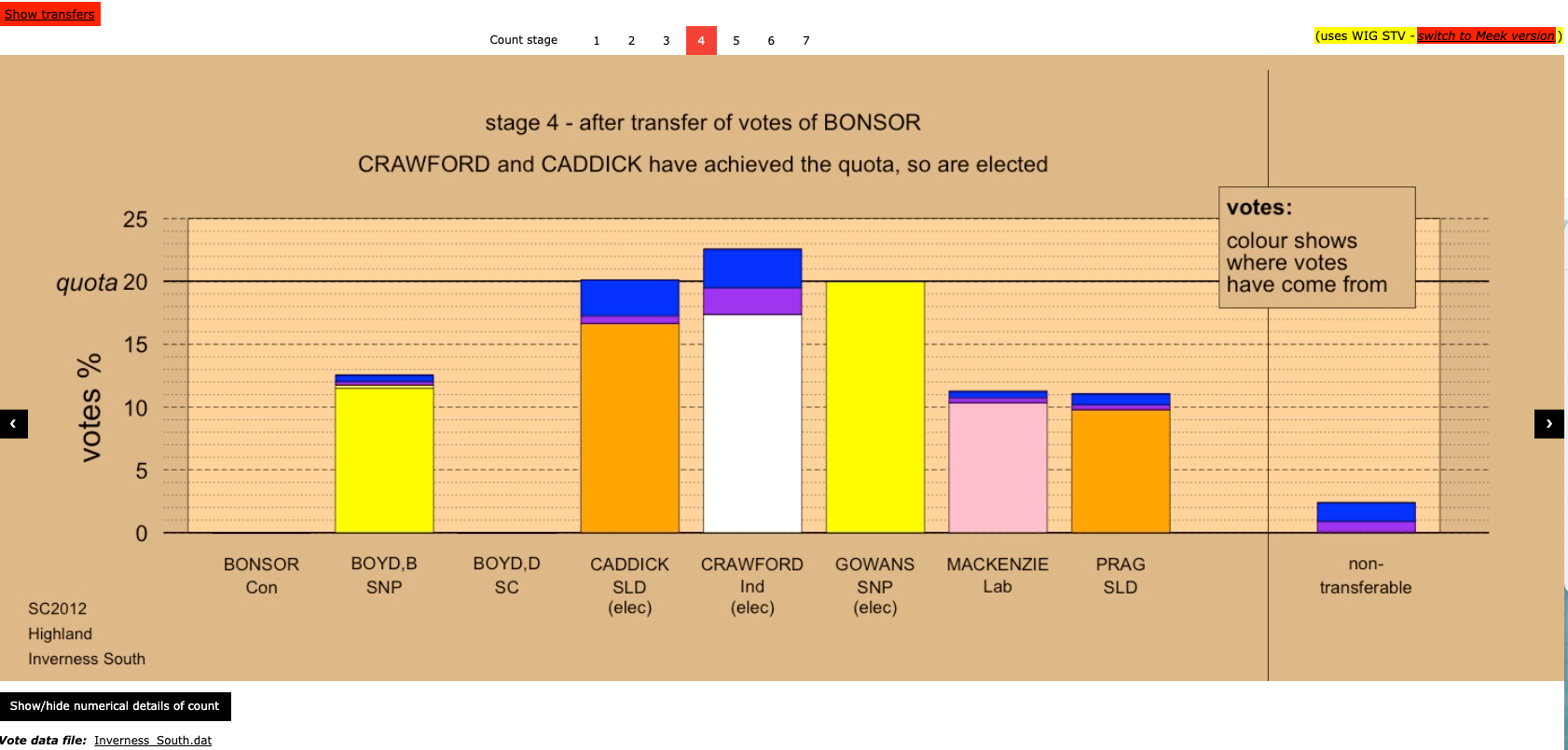}}
\noindent
\caption{\label{stv_stage}Example of graphic output for one stage of a Scottish Council election}
\end{figure}

One of List-PR's advantages is that its results are very straightforward to present.
MMP is much more of a problem, especially if one wishes to show how near or far the vote distribution is from having disproportional overhangs.
The Supplementary material includes \href{https://www.macs.hw.ac.uk/~denis/mmp_elections/scotland2021.html}{an attempt to visualise the results}, identifying crucial contests, for the Scottish Parliament election of 2021.

\section{Discussion}
\label{discussion}

\subsection{System choice}
\label{system}

This paper has evaluated the main proportional representation systems against the `good systems' criteria discussed in \S\ref{criteria}, principally voter choice, proportionality and local representation.
And for the key criterion of proportionality, it has gone beyond using one simple statistic to considering the three aspects that matter most for fair representation - linearity, and thresholds for representation and for a majority.

It remains to summarise the relative advantages of the three main systems considered.
Because their fit to criteria depends significantly on how they are implemented, particularly on districting and certain system-specific rules, it is important for a fair comparison that we consider best-practice versions of each system.

 \subsection*{STV}
  
For STV, we have found that there are clear advantages in using variable-sized districts based on natural community boundaries, with sizes  in the range either 3-5 or 4-7 seats, but allowing some flexibility, especially for isolated or sparse areas.
As to detailed rules, see the discussions in {\S}s \ref{history}, \ref{choice} and \ref{problems}.

Because STV is defined in terms of voter choice, it is no surprise that it comes out best on that criterion.
Nor is it surprising that it gives good local representation, as judged by the proportion of voters getting a representative they voted for, because the concept of STV is based on making a best choice of representatives (\S\ref{stv}).
What is perhaps surprising is that STV - whether with fixed or variable district size - comes out best for all three aspects of proportionality, even though only first preferences are considered in this comparison.
The explanation lies in the transfer system of STV, which avoids wasted votes and gives a smoother threshold for representation ({\em e.g.} Figure \ref{thresh_ni}).
The latter, meaning greater uncertainty in how many first preference votes are required for election, is perhaps a downside for the individual candidate, but seems to be intrinsic to getting better proportionality for parties overall.

Perhaps the most novel, and certainly the most satisfying, piece of analysis of this paper has been the fresh look at the problem of districting for STV.
The approach of fitting electoral districts to communities  without worrying about parity has probably been avoided in the past because it is assumed that it must come with significant downsides, particularly as regards proportionality.
But it turns out to have almost exclusively upsides, including being better for proportionality in several ways.
Perhaps its most important advantage is easy updating (\S\ref{local}), which avoids the biases in representation that arise when, as is the norm with traditional boundary reassessments, elections take place with allocations that are many years out of date (\citep{Baston:2010}).

The variation in size that is necessary for the natural communities approach also turns out to be an advantage in most respects.
It is likely to provide a better fit to the national vote distribution because different sizes have different cutoff levels (\S\ref{equalisation}, Figure \ref{brexit}).
Further, the use of natural rather than arbitrary boundaries can be expected to give greater variation in vote distribution between districts, which in turn means that the discretisation approximation to that distribution provided by  STV is likely to be more accurate, while at the same time giving better opportunities to minor parties and locally grounded candidates.
The variation in threshold for representation (\S\ref{thresh}) that goes with variation in size - and is larger for STV in any case (Figure \ref{thresh_ni}) - means that for an individual the probability of election as a function of first preference votes increases from 0 to 1 over a wide range, as opposed to the rather sharp - almost deterministic - cutoff of list systems.

 \subsection*{List\_PR}
 
Most of the advantages for local representation in using natural community boundaries (including more meaningful representation and better proportionality because easier to update) apply equally to List-PR and MMP.
Indeed, such use of natural boundaries is constitutionally enshrined in many electoral systems, for example those of the Nordic countries (List-PR) and Germany (MMP).
Where such mandated boundaries do not exist, the size of districts to choose will need to trade off the desiderata of localism and proportionality.
For both criteria, detailed consideration suggests a slightly higher number than for STV, so perhaps a maximum of around 10 for List-PR districts.
Using St-Lague with constituencies electing 6-10 representatives should ensure good proportionality between major parties, with a threshold for representation of at most about 5\% of the national vote, and considerably lower for parties with local concentrations of support.
All constituencies will be competitive, so that votes matter equally everywhere.

`Good overall proportionality' here is of course ignoring the wasted vote problem of all single-preference systems, whereby votes for below-threshold parties contribute nothing to results, a problem compounded by the deterrent this implies against expressing real preferences - which in turn threatens the concept of proportionality.

The key criterion on which List-PR and MMP fall down is voter choice: voters should be able to express support for a large or small party or none without fear that their vote will be wasted, and to express preference for individual candidates within parties..
These problems can be mitigated to some extent by (respectively) allowing parties to present combined lists, and by the use of open lists; the electoral system of Finland includes both these features.

In sum, if the great majority of voters have strong party allegiances, and  are happy to vote for a party rather than for individuals, then List-PR is a reasonably satisfactory system.

\subsection*{MMP}

For districting in MMP, similar considerations apply, except that somewhat larger districts are appropriate.
For example, a district with 9 constituencies and 9 list seats will give voters the same number of representatives as a List-PR district of size 10.
Having such larger districts should also somewhat reduce the probability of occurrence of overhangs.
However, the use of more politically coherent districts, which tends to improve the proportionality of 
STV and List-PR, works against MMP because coherent districts are more likely to have overhangs.

MMP retains the single local representative that some find attractive, but at the expense of major problems: these include safe seats, having two different kinds of representatives, and difficulty for voters in understanding the effect of their votes (\S\ref{mmp2}).
However, the biggest problems with  MMP are the potential for overhangs, with their nonlinear effect on proportionality (\S\ref{majority}), and for gaming the system (\S\ref{mmp2}).
Neither STV nor List-PR has such problems.

\subsection*{In conclusion}
To summarise with a broad brush: FPTP, alias plurality, fails on on all three of the basic criteria.
Its strongest suit is in providing very local representation, but the costs of this include non-natural boundaries needing frequent adjustment; around half of voters not having a representative they voted for; and probably worst, some voters - swing voters in marginal constituencies - have much more influence than others, thus significantly distorting politics.

List-PR addresses all three of these problems, with votes mattering equally everywhere, and introduces the major improvement in fairness of broad proportionality between votes and seats.

MMP is a halfway step from FPTP towards List-PR, a compromise that brings with it the problems summarised above.

STV on the other hand can be seen as a refinement of List-PR (see \S\ref{list}), in which allowing voters to more fully express their preferences and allowing transfers so as to avoid wasted votes significantly improves the proportionality of the system as well as voter choice; and can be implemented at the smallest local scale that gives most voters a representative that they voted for.


\subsection{Technical challenges}
\label{technical}

The statistical analysis of this paper has concentrated on the socially most important questions of systems analysis in relation to basic democratic criteria.
But as a coda, here are a few problem areas of interest to the theoretician - and like much theory, progress in these areas may lead in time to socially useful conclusions.

The body of nearly 1100 sets of full preference data for Scotland's recent council elections is a huge resource, whose surface I have only scratched here.
The dependence of second on first preferences has been touched on briefly in \S\ref{patterns}; a rather different approach is taken in a parallel paper on electing a single choice \citep{Mollison:2023}, where the occurrence or (spoiler alert) mainly non-occurrence of cyclic preferences is of key interest.

In a multi-party context, such as the Scottish Council elections where there are five principal parties and the relations between them have changed over successive elections (Figure 9), it would be interesting to look deeper into the structure of preferences.

Finally, in the analysis of the vital threshold for a majority (\S\ref{majority_stv}) it was pointed out that the support for a potential majority is not just a matter of first preferences; under STV it depends on the balance of final preferences.
The question of how first and final preferences are related, and what it tells us about the relations between parties and the voters' perceptions of these, is of significant social interest.

\bigskip

\section*{Supplementary material}
\label{supp}

\begin{enumerate}
\item[\S\ref{local}] \href{https://www.macs.hw.ac.uk/~denis/STV2023.pdf}{An STV scheme for Westminster}
\item["] \href{https://www.macs.hw.ac.uk/~denis/stv/futurereviews.pdf}{The Islands Review proposals 2021 and their implications}
\item[\S\ref{prop}] Data, R code and additional figures for proportionality analysis {\em (to come)}
\item[\S\ref{patterns}] Data, R code and additional figures for preference analyses {\em (to come)}
\item[\S\ref{problems}] \href{https://www.macs.hw.ac.uk/~denis/stv/fine-tuningSTV.pdf}{Community-centred democracy: fine-tuning the STV Council election system}
\item[\S\ref{presentation}] \href{https://www.macs.hw.ac.uk/~denis/stv_pkg_manual.pdf}{pref R package manual}
\item["] \href{https://www.macs.hw.ac.uk/~denis/mmp_elections/scotland2021.html}{Results of the Scottish Parliament election of 2021}
\end{enumerate}

\bigskip

\bibliography{fair}

\end{document}